# Realization of Associative Memory in an Enzymatic Process: Towards Biomolecular Networks with Learning and Unlearning Functionalities

Vera Bocharova,[1] Kevin MacVittie,[1] Soujanya Chinnapareddy,[1]

Jan Halámek,[1] Vladimir Privman[2] and Evgeny Katz[1]

[1] Department of Chemistry and Biomolecular Science, [2] Department of Physics, Clarkson University, Potsdam, NY 13699, USA.

ABSTRACT: We report a realization of an associative memory signal/information processing system based on simple enzyme-catalyzed biochemical reactions. Optically detected chemical output is always obtained in response to the triggering input, but the system can also "learn" by association, to later respond to the second input if it is initially applied in combination with the triggering input as the "training" step. This second chemical input is not self-reinforcing in the present system, which therefore can later "unlearn" to react to the second input if it is applied several times on its own. Such processing steps realized with (bio)chemical kinetics promise applications of bio-inspired/memory-involving components in "networked" (concatenated) biomolecular processes for multi-signal sensing and complex information processing.

# TOC Graphic

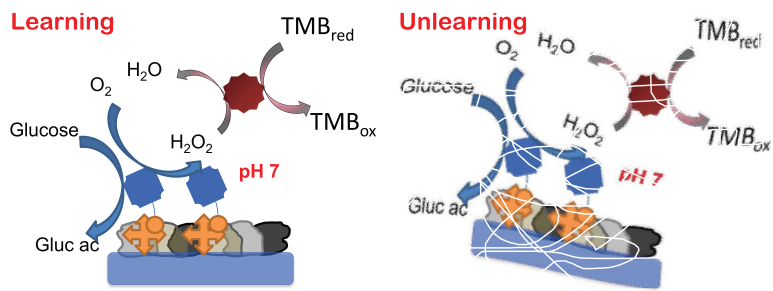

**Keywords:** Associative memory, Biocomputing, Enzyme, Biomolecular network, Biomolecular information processing, Unconventional computing

SECTION: Biophysical Chemistry

---

Recently, there has been growing interest in novel device-component and network-element development[1] for information and signal processing based on approaches alternative to conventional electronics.[2,3] This research has been driven by promise of new functionalities in computing, interfacing, and multi-input sensing using molecular systems.[4-9] Specifically, biomolecular information processing[10] has offered a promising approach allowing biocompatibility and ability to utilize naturally available and bio-inspired/derived synthetic molecules with specificity and selectivity for tasks in signal and information processing.[11-12]

The paradigm for devising information processing steps has been in most cases that of binary gates and their networking, attempting to mimic the well-developed digital approach of modern electronic circuitry. Indeed, the near-future biomolecular information processing approaches are unlikely to offer direct competition to the speed and versatility of modern computers. Therefore, the emphasis has been on additional functionalities and interfacing capabilities aimed at supplementing electronic systems. Various binary gates, such as AND, OR, XOR, etc., have been realized[1,10-12] in biochemical processes involving DNA/RNA,[13-15] proteins/enzymes,[12,16] and even whole cells.[17,18] For enzyme-process based systems considered here, preliminary few-step networks designed[12,19] for specific sensing and diagnostic applications[20] have been reported. Noise-handling properties of such networks and optimization of their constituent gates have been explored.[21]

As an interesting alternative to the "binary gate" approach to network elements, utilization of paradigms borrowed from natural processes can prove beneficial, especially for information processing with biomolecules. Such ideas warrant investigation notwithstanding the fact that the full understanding of complex information processing in natural processes is not yet available.[22] Memory properties are the most basic features of functioning of natural processes, systems and organisms on all scales. Not surprisingly, there has generally been recent interest in realizing device components involving memory features, exemplified by memristors[23-26] and other memory-involving designs.[27-30] These have primarily been accomplished with electronic devices rather than in biomolecular systems.

Here we report the first realization of a simple variant of associative memory in an enzymatic biochemical process. Generally, the concept of associative memory is complicated and diverse, and various definitions of it are possible. It has traditionally been studied in systems in biology and computer science. For instance, some such memory/learning properties have been extensively explored in neural networks.[30] However, here we consider memory features directly based on physical or chemical processes. We demonstrate "learning" to respond to **Input 2** after it was applied together with the main, response-triggering **Input 1**. In terms of chemical kinetics this means that the system actually restructures in such a way that not only the original input but also the second one can initiate the process of producing the response signal. Furthermore, if **Input 1** is removed, then the system "forgets" (unlearns) it's training and no longer responds to **Input 2** after several applications of the latter. The devised biochemical-reaction process is simple enough to be a candidate for connecting with other biochemical "gates" and network components. This research offers interesting possibilities of incorporating bio-inspired, memory elements in designs aimed at increasing the complexity of biomolecular-computing systems.

Our system is outlined in Scheme 1 (see the next page), where we show the "learning" steps. Both "learning" and "unlearning" are described in terms of biocatalytic processes below. Biocatalytic reactions were realized using glucose oxidase (GOx) from *Aspergillus niger* type X-S (E.C. 1.1.3.4), horseradish peroxidase (HRP) type VI (E.C. 1.11.1.7), urease from *Canavalia ensiformis* (jack bean) (E.C. 3.5.1.5) and iminobiotin-GOx conjugates (the full list of chemicals used, and the procedure for preparation of iminobiotin-GOx conjugates are detailed in the *Supporting Information*, appended as the last two pages to this preprint). 96-well polystyrene microtiter ELISA plates (VWR) were functioning as processing reservoirs. 100 μL of 10 μg/mL concentration of avidin in 0.1 M carbonate buffer, pH 9.6 was layered on the surface of the well by simple physical adsorption at 4 °C overnight. The excess of avidin was removed by washing each well four times with 300 μL of 100 mM phosphate buffer saline (PBS) with 0.05% (v/v) Tween-20 (PBST) pH 7.4. The blocking step (to prevent further nonspecific adsorption of proteins) was accomplished via the addition of 300 μL of a blocking solution (10 mg/mL bovine serum albumin (BSA) dissolved in PBS) to each avidin-coated well. The excess of BSA was removed after 1 h of incubation at 25 °C by washing each well four times with PBST, yielding the avidin-functionalized surface.

The following ingredients were applied as input signals: **Input 1** consisted of 30 μM of $H_2O_2$ and 0.2 mg/mL of iminobiotin-GOx conjugates. **Input 2** consisted of 1 mM glucose and 5 mM urea. All the inputs were dissolved in the active media composed of 5 mM citrate buffer pH 4.5, containing 83 μM 3,3′,5,5′-tetramethylbenzidine (TMB), 1 mU/mL HRP and 1 U/mL urease. The final volume placed on the plate was adjusted to 100 μL. Wells containing all the inputs but enzymes, served as controls. In order to check the stability of GOx-conjugates under ambient conditions, the plate was incubated with a combination of **Input 1** and **Input 2** for half an hour. The unattached residuals were washed away according to the aforementioned procedure with PBST and **Input 2** was applied to each row with interval of 30 min. Multiple signals were applied to a single digital processing reservoir, but not exceeding more than 4 applications, as the signal seemed to deteriorate. The output signal of oxidized TMB was read out in 30 min at 655 nm using a BIO RAD Model 680 ELISA reader.

**Scheme 1. Schematic representation of the "learning" processes in an ELISA well.**[a]

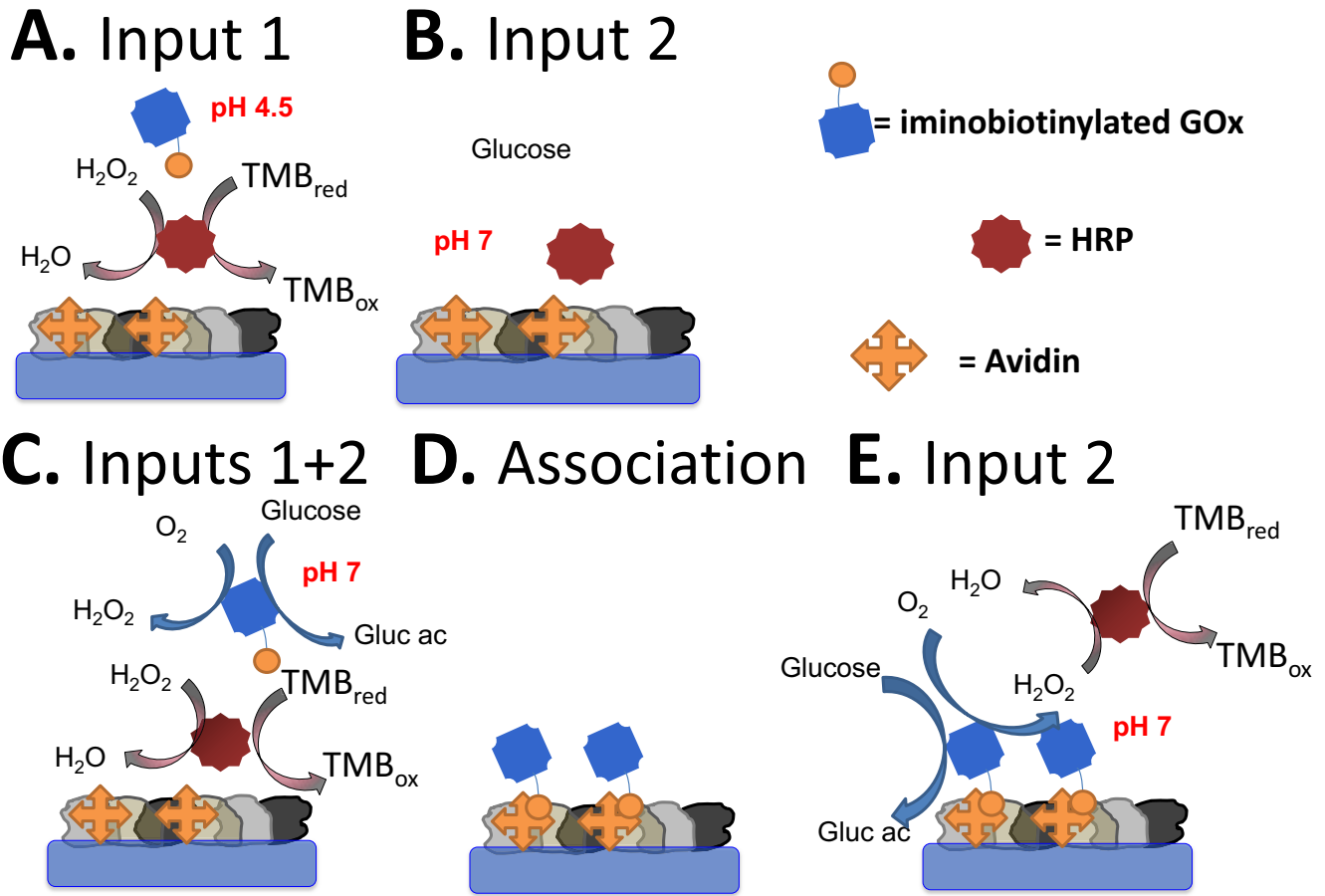

[a] (A) **Input 1** only: combination of $H_2O_2$ and iminobiotin-GOx conjugates, dissolved in active media consisting of mixture of HRP, TMB and urease in citrate buffer at pH 4.5. The output signal is measured as the oxidation of TMB. (B) **Input 2** only: combination of glucose and urea in the same active media. No output signal is generated. (C) Simultaneous introduction of **Inputs 1** and **2**. The output signal is present. (D) "Learning" upon application of both inputs. The surface of the well is left with the iminobiotin-GOx conjugates attached to avidin. (E) Application of **Input 2** after the "learning" step, results in an output signal. Note that not all the components are shown in the panels, and that pH 7 is a result of the simultaneous presence of urease and urea.

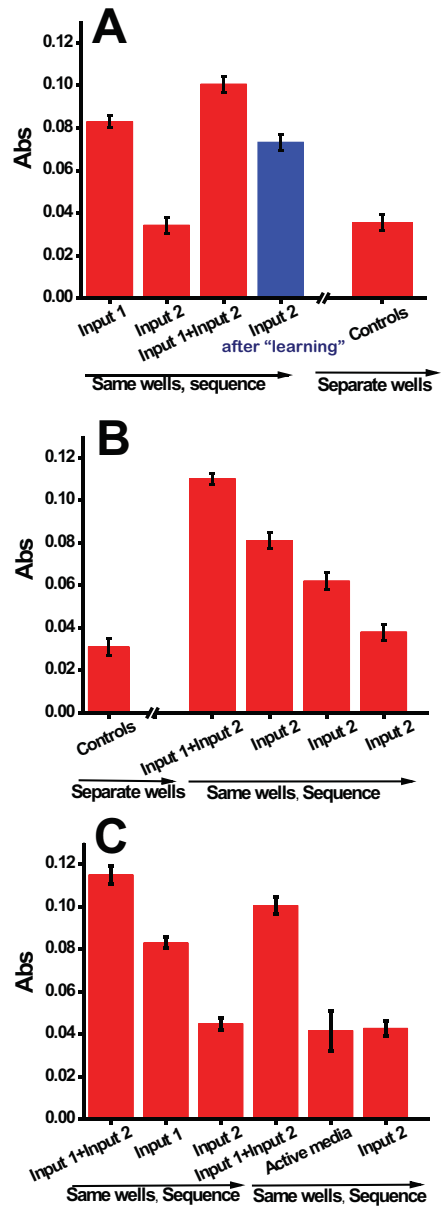

**Figure 1.** (A) The "learning" steps. The red bars show typical optical signals for **Input 1** and **Input 2** when applied to the same ELISA wells of the initial system separately or simultaneously. The blue bar shows the optical signal in response to **Input 2**, after the system "learned" from an earlier application of **Input 1** and **Input 2** together. (B) Repeated application of **Input 2** to the same wells, after **Input 1** and **Input 2** were first applied simultaneously, results in unlearning. (C) Reset ("rapid forgetting"), when **Input 1** or active media were applied right after **Input 1** and **Input 2** were applied together. Notes: The active media were renewed for each signal or signal combination application. The value of the background noise for (C) is the same as for (B). The signal was generated as the oxidized TMB and detected by ELISA reader after 30 min at 655 nm.

We have demonstrated associative memory in a simple enzymatic network consisting of two input signals dissolved in active media applied separately or simultaneously to ELISA plate wells modified with avidin. As outlined in Scheme 1, **Input 1** is a combination of $H_2O_2$ and iminobiotin-GOx conjugates in the active media (mixture of HRP, TMB and urease in the citrate buffer of pH 4.5). The system generates an output signal in response to **Input 1**. When **Input 1** is applied, TMB is oxidized by HRP enzyme in the presence of $H_2O_2$ substrate. The recognition response for the present system is thus the optical signal monitored at 655 nm after 30 min of the reaction. Typical intensity of the output signal for **Input 1** is shown in Figure 1A.

**Input 2** is the combination of glucose and urea in the active media. The original system is not responsive to **Input 2**, and no signal is present beyond the background, see Figure 1A. However, the system can be "taught" to recognize **Input 2** and associate it with the response characteristic of **Input 1**. In order to tie **Input 2** to the specific recognition reaction, **Inputs 1** and **2** were first applied together. At the chemical level, the kinetic step of "learning" is related to the attachment of iminobiotin-GOx conjugates to the surface of the avidin-functionalized ELISA plate. The binding occurs at basic pH, whereas elution can be accomplished by changing the pH to acidic. We found that pH 7 is an optimal compromise pH value to achieve a weak binding[31] of iminobiotin-GOx conjugates to the avidin-functionalized surface of the wells, while keeping our enzymes active; see Figure SI1 in the *Supporting Information*.

The "learning" process requires some time to achieve the required pH change. Indeed, the starting pH of the reaction in our system was 4.5. However, when urea of **Input 2** and urease of the active media react, the pH is changed, see Figure 2A (on the next page), causing partial deprotonation of iminobiotin and attachment of the iminobiotin-GOx conjugates of **Input 1** to the avidin-modified surface of the ELISA plate. The time of the reaction should be selected large enough for the pH to stabilize. Furthermore, it is desirable to have the output signal at approximately the same values whenever there is a response. Our preliminary work with the present setup suggested that 30 min is a good selection; cf.

Figure 2B. It should be noted that simultaneous application of **Input 1** and **Input 2** resulted in pH changes which are favorable for the formation of completely oxidized form of TMB, thus converting a fraction of TMB to the second oxidation state absorbing at 450 nm. Thus, at first the oxidation process resulted in the absorbance increase at 655 nm, corresponding to the formation of a radical cation that forms a charge-transfer complex with the unoxidized compound (denoted as $TMB_{ox}$ in Scheme 1). Later, the completely oxidized form (diimine) begins to form yielding the absorbance decrease. Application of **Input 2** separately (after the learning process) resulted in the formation of the first oxidation state of TMB without its overoxidation, thus demonstrating only the increasing (with time) absorbance at 655 nm. The system was optimized to have similar optical output signals in both cases at reaction times of 30 min.

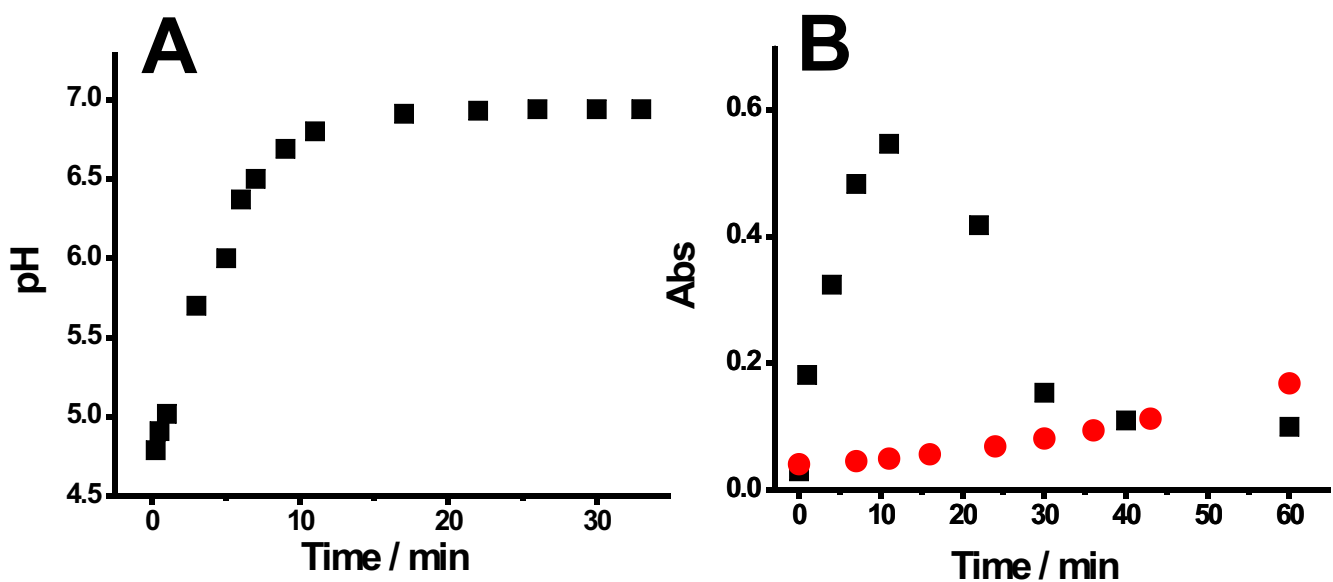

**Figure 2.** The square symbols denote (A) variation of the pH, and (B) absorbance (at $\lambda$ = 655 nm), when **Inputs 1** and **2** were applied together. The same pH variation was obtained when **Input 2** was applied separately. The circle symbols in panel (B) represent the absorbance for **Input 2** following the "learning" step of a simultaneous application of **Inputs 1** and **2**.

When **Input 2** is applied following a simultaneous application of **Input 1** and **Input 2**, the system will "learn" to recognize **Input 2** as it does **Input 1** (Figure 1A). This occurs because the adsorbed GOx uses glucose as its substrate and produces $H_2O_2$ *in situ*. $H_2O_2$ is then reduced by HRP, while TMB is oxidized, producing the output signal. The system thus "learned" to respond to **Input 2**. Figure SI2 in the *Supporting Information* illustrates that the intensity of the recognition reaction did not change even after the plate was exposed to ambient conditions for 2 hours, thus suggesting that the system can "memorize" its state for a long time, once "trained" by the combined application of **Inputs 1** and **2**.

The "unlearning" step occurs after a repeated application (each for 30 min) of **Input 2** to the same wells, which were first pre-treated ("trained") with a single 30 min application of **Inputs 1** and **2** together. This leads to a gradual decrease of the output signal, see Figure 1B, due to partial detachment of the iminobiotin-GOx conjugates from the plate. The initial pH is acidic and this causes the onset of detachment of the conjugates. However, as the urea of **Input 2** and urease of the active media change the pH to basic, the conjugates became deprotonated and get reattached to the surface. However, only a fraction of the conjugates is reattached because the solution with the conjugates is less concentrated in each step due to the replacement of the active media. The extent of "unlearning" can be quantified by observing the decreasing system responses, Figure 2B.

"Rapid forgetting" (reset) of the system can be achieved if only **Input 1** or just the active media are applied to the system right after **Inputs 1** and **2** were applied simultaneously, see Figure 1C, because this causes detachment of conjugates at acidic pH.

In conclusion, our present results demonstrate that, bio-inspired processing steps with memory, can be realized with simple enzymatic processes. Future work should not only address design and optimization of such systems, as well as other memory-involving network elements of interest, but also their incorporation in more complicated "networks" for sensing and processing of information.

ACKNOWLEDGMENT This research was supported by the National Science Foundation (Awards CCF-1015983 and CBET-1066397).

**Supporting Information Available:** List of chemicals; preparation of iminobiotin-GOx conjugates; and additional figures showing activity of iminobiotin-GOx conjugates measured at different pH values and as a function of time. *Supporting Information is appended as the last two pages of this preprint.*

# Realization of Associative Memory in an Enzymatic Process: Towards Biomolecular Networks with Learning and Unlearning Functionalities

Vera Bocharova,[1] Kevin MacVittie,[1] Soujanya Chinnapareddy,[1]
Jan Halámek,[1] Vladimir Privman[2] and Evgeny Katz[1]

[1] Department of Chemistry and Biomolecular Science, [2] Department of Physics,
Clarkson University, Potsdam, NY 13699, USA

## Supporting Information

**Chemicals and reagents.** Glucose oxidase (GOx) from *Aspergillus niger* type X-S (E.C. 1.1.3.4), horseradish peroxidase (HRP) type VI (E.C. 1.11.1.7), urease from *Canavalia ensiformis* (jack bean) (E.C. 3.5.1.5), avidin from egg white, bovine serum albumin (BSA), urea, β-D-(+)-glucose, 3,3′,5,5′-tetramethylbenzidine (TMB) were purchased from Sigma-Aldrich. Trifluoroacetamidoiminobiotin-NHS ester was purchased from Marker Gene Technologies Inc. Hydrogen peroxide 30% was purchased from J. T. Baker. All commercial chemicals were used as supplied without further purification. The trifluoroacetamidoiminobiotin-GOx (iminobiotin-GOx) conjugate was prepared via the coupling procedure described below. Ultrapure water (18.2 MΩ·cm) from a NANOpure Diamond (Barnstead) source was used in all of the experiments.

**Preparation of iminobiotin-GOx conjugates.** A 1 mL solution of 0.31 μmol GOx and 8.26 μmol trifluoroacetamidoiminobiotin-NHS ester in 100 mM phosphate buffer saline (PBS), pH 7.4, was prepared. The molar ratio of GOx and trifluoroacetamidoiminobiotin-NHS ester was approx. 1:26. After three hours of incubation at 4 °C, the solution was transferred to a 500 μL centrifugal filter tube with a molecular weight cut-off equal to 30 kDa and centrifuged at 12,000 rpm for 5 minutes. The procedure was repeated 4 times to ensure the filtration of unbound trifluoroacetamidoiminobiotin. The final volume was adjusted to 500 μL and concentration of the conjugate was 100 mg/mL. Finally, aliquots were made (20 μL in each tube). Aliquots were unfrozen before use and each aliquot was dissolved in 10 mL of a corresponding buffer solution with the final concentration of enzyme of 0.2 mg/mL.

**Additional figures.**

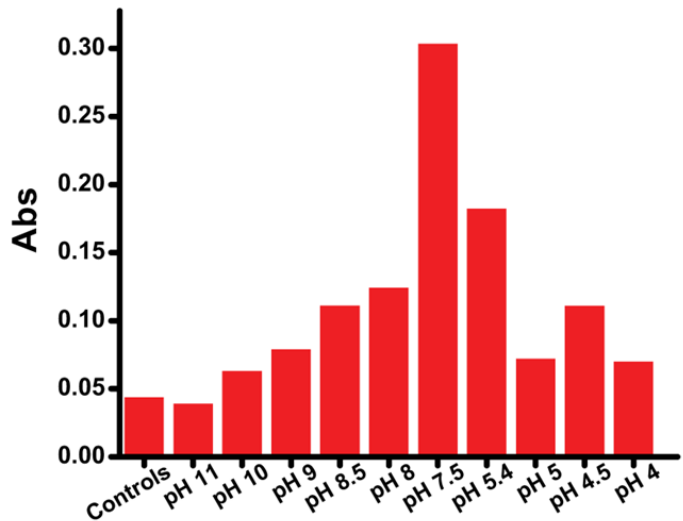

**Figure SI1.** Activity of iminobiotin-GOx conjugates adsorbed at the surface of ELISA plate modified with avidin from solution at different pH. The "control" bar here is the average from the highest values for solutions with substrates, but without the conjugates, at different pH values.

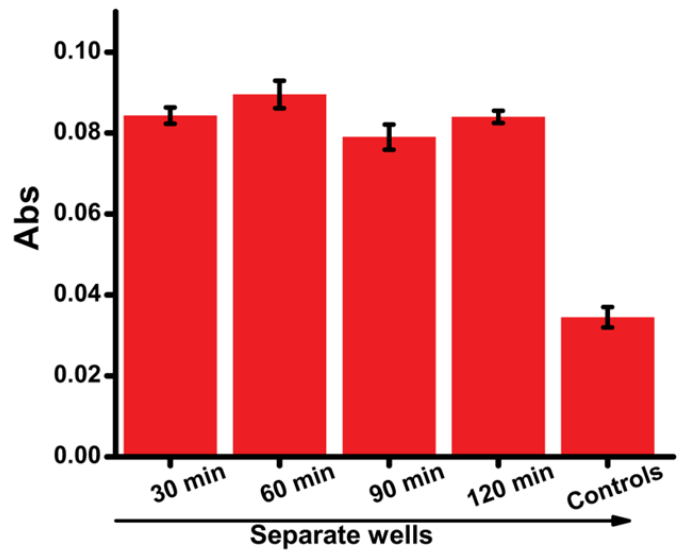

**Figure SI2.** Activity of iminobiotin-GOx conjugates exposed to ambient condition. Measurements were performed in different wells of ELISA plate. To achieve statistical significance, each experiment was done in a row of 8 wells.